# The Complex Systems and Biomedical Sciences group at the ESRF: current status and new opportunities after Extremely Brilliant Source upgrade.

Maciej Jankowski,[1*] Valentina Belova,[1,2] Yuriy Chushkin[1], Federico Zontone[1], Matteo Levantino,[1] Theyencheri Narayanan[1], Oleg Konovalov,[1] Annalisa Pastore[1]

[1]The European Synchrotron- ESRF, 71 Avenue des Martyrs, CS 40220, 38043 Grenoble Cedex 9, France

[2] Univ. Grenoble Alpes, CEA, IRIG/MEM/NRS 38000 Grenoble, France

Correspondence e-mail: maciej.jankowski@esrf.fr

## Abstract

The Complex System and Biomedical Sciences (CBS) group at the European Synchrotron Radiation Facility (ESRF) in Grenoble is dedicated to the study of a broad family of materials and systems, including soft and hard condensed matter, nanomaterials, and biological materials. The main experimental methods used for this purpose are X-ray diffraction, reflectivity, scattering, photon correlation spectroscopy, and time-resolved X-ray scattering/diffraction. After a recent and successful Extremely Brilliant Source (EBS) upgrade, the Grenoble synchrotron has become the first of the 4[th] generation high energy facilities, which offers unprecedented beam parameters for its user community, bringing new experimental opportunities for the exploration of the nanoscale structure, kinetics, and dynamics of a myriad of systems. In this contribution, we present the impact of the recent upgrade on the selected beamlines in the CBS group and a summary of recent scientific activities after the facility reopening.

## 1. Introduction

Over the last decades, the rapid development in experimental instrumentation and theory allowed studies of various families of materials and molecules using X-ray radiation. The construction of modern synchrotron facilities has, for instance, enabled a plethora of experiments concentrating on the structural investigations of condensed matter, often in real-time, *in situ*, in *operando*, and in sample environments not accessible to other experimental techniques [1], e.g., electron-based or scanning probe microscopy. X-ray diffraction and scattering are among the most utilized methods at synchrotrons, providing powerful possibilities to gain insight into the structure of matter at scales ranging from sub-Ångstroms to micrometers [2]. In addition, being employed in a time-resolved mode, these methods allow investigations of kinetics [3] and dynamics [4] of the studied systems, with a time resolution reaching up to hundreds of picoseconds [5].

Among the synchrotron radiation institutions is the European Synchrotron (ESRF), an international research facility supported by 22 countries and based in Grenoble, France. Thanks to high-level, innovative engineering and cutting-edge vision, the ESRF is a recognized leader in its field. It typically hosts several thousand scientists every year, from all over the world, working both in academia and in industry, to study the inner structure of materials and living matter down to atomic resolution. This European facility was founded in 1988 and began operation in 1994. Since then, ESRF has had regular upgrades to keep up with the technology. The most recent upgrade accommodated the Extremely Brilliant Source (EBS) project, which has brought new exciting opportunities to its user community [6]. The upgrade to the 4$^{th}$ generation synchrotron was started in autumn 2018 and finalized in 2020, with an opening to the user community in August 2020, despite the COVID-SARS-19 pandemic and related restrictions. The storage ring lattice was entirely replaced by a novel hybrid multi-bend achromat (HMBA) lattice, reducing the beam's horizontal emittance and increasing brilliance and the coherent flux of generated photons by a 100 times factor.

The more than 30 beamlines at ESRF are clustered into six groups. The CBS group is composed of beamlines utilizing X-ray diffraction and scattering as their main experimental techniques for investigations of polymers, gels, colloids, detergents, nanomaterials, proteins, two-dimensional (2D) materials, membranes, thin films, liquids, and glasses. Among the CBS beamlines, the ID02 beamline is specialized in the application of X-ray scattering methods, especially to study biological systems and complex soft matter, offering a high spatial and temporal resolution in the reciprocal space and employing coherent X-ray scattering [3]. The ID09 beamline uses a pink beam and a dedicated X-ray chopper to study fast structural changes in materials exposed to short laser pulses, i.e., pump-probe experiments [5]. Finally, the ID10 beamline comprises two endstations. The first endstation is dedicated to the studies of liquid and solid interfaces and surfaces, whereas the second endstation focuses on matter dynamics and high-resolution imaging probed by coherent X-ray scattering [3]. All these beamlines allow the study of the complex soft matter and related phenomena like phase transition [3], self-assembly [3,7,8], virus capsid assembly [9,10], muscle regulation [11,12], protein folding [13], and protein dynamics [14]. Moreover, the expertise of the beamlines covers some fields of hard condensed matter, i.e., 3D imagining [15], thin films [16], perovskites [17,18], 2D materials [19], microelectronics [20], molecular dynamics [21] and industrial research [22]. These examples show the broad scientific portfolio available for the user community at the CBS beamlines, soon complemented by the new 200 m long beamline dedicated to coherent X-ray scattering, currently under construction at the ID18 site [23].

The EBS implementation has enormously benefitted the ESRF beamlines using coherent X-ray scattering, i.e., ID10 and ID02, which can now access structural information of materials hardly achievable before the upgrade [3] or construction of beamlines exploiting the new possibilities [23]. Furthermore, besides the increase in the coherent flux, the ESRF beam source size in the horizontal direction has decreased by almost 30 times, which can be helpful in investigations where sample scanning can bring additional information [24]. Overall, the total flux of the X-ray beam has increased by several folds, which is vital for the pump-probe experiments at ID09, or for studies of 2D materials with a low scattering cross-section at ID10 [19].

In this contribution, we present the characteristics of three selected CBS beamlines that utilize X-ray diffraction and scattering as primary techniques. First, the impact of the EBS upgrade is quantitatively presented for the ID10 beamline, where the experimental measurements and theoretical calculations show in detail the actual gain from the EBS. Next, the impact of EBS on the time-resolved experiments is described for the ID10 coherent X-ray scattering endstation, ID09, and ID02 parts. Finally, the current research trends and topics are described for each beamline, presenting the complete portfolio of the CBS group capabilities.

## 2. ID10 Soft Interfaces and Coherent Scattering Beamline

The ID10 beamline comprises of two endstations that share the same X-ray optics. The first endstation is dedicated to investigations of liquid and solid surfaces using wide-angle X-ray scattering (WAXS), small-angle X-ray scattering (SAXS), and X-ray reflectivity (XRR). The second endstation is assigned to study matter using X-ray photon correlation spectroscopy (XPCS) and coherent X-ray scattering. In brief, the beamline uses a high-brilliance X-ray beam formed by three undulators, covering an energy range from 7 to 30 keV. The double-mirror setup at the beginning of the optical hutch strongly suppresses the higher harmonics of the beam. Consequently, a liquid nitrogen-cooled channel-cut crystal monochromator, composed of Si(111) and Si(311) crystals used alternately, is employed for beam monochromatization with the resolution of $\Delta E/E$: $1.4\times10^{-4}$ and $2.7\times10^{-5}$, respectively. Finally, the set of refractive lenses focuses the X-ray beam at the sample position to the size between a few and a few tens micrometers, depending on the specific optics settings and distance from the source. It is worth mentioning that the recent EBS upgrade and beamline monochromator upgrade resulted in a fourfold gain in beam flux at the sample position, reaching $1\times10^{13}$ ph/s at high energies, above 20 keV.

To illustrate in more detail the impact of EBS on the beam parameters, we compared the spectral flux measured experimentally and simulated. Fig. 1 shows the spectral flux through the primary slits (located at 27.2 m from the source) open at 0.15 x 0.15 mm of the undulator U27 tuned with the first harmonic at 7.3 keV with EBS (black circle)s and at 8 keV (blue circles) with the old high $\beta$ source over the first three harmonics (Fig. 1A ) was compared with the corresponding brilliance after normalization to the photon beam parameters (Fig. 1B). The spectral flux was measured by scanning the Si(111) channel-cut monochromator and counting the photons scattered by a thin Kapton foil in 90° vertical scattering geometry [25]. The measured count rates in counts/sec were converted into flux in photons/s/0.1%bw, considering the transmission of the beamline and the bandwidth of the monochromator. The curves with a continuous line show the theoretical computation of the synchrotron radiation emission generated by the software package SRW, which includes the phases space properties of the electron beam at the exact location of the insertion device and its energy spread [26] and match quite well the measured flux over the entire scanned energy range.

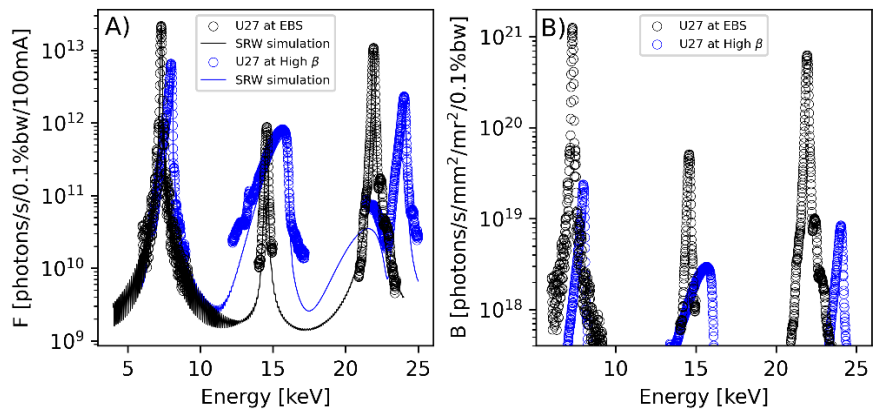

Fig. 1. Comparison of the spectral flux measured experimentally and simulated. A) the measured spectral flux F at EBS compared to the old "high-β" electron source. Curves in solid lines are simulations with the software package SRW according to the electron beam parameters at the source location and energy spread. B) the corresponding brilliance after normalization, the photon beam source size, source divergence, and bandwidth.

The spectral flux at EBS through a small slit is increased thanks to the higher collimation of the beam and the narrower spectral shape of the harmonics due to the smaller emittance of the electron beam. Let us note that this gain will be less and less pronounced at larger slit apertures since the total emission of the insertion device did not change with EBS, as the accelerator parameters have been kept unchanged, e.g., the energy at 6 GeV. However, the drop in (horizontal) emittance from 4000 pm to 130 pm greatly impacts the brilliance with a gain of almost two orders of magnitude.

## 2.1. ID10 coherent X-ray scattering station

The ID10 coherent X-ray scattering is the endstation of the ID10 beamline, located at about 60 m from the source point. It is dedicated to X-ray scattering with a coherent X-ray beam, using XPCS and tomographic Coherent X-ray Diffraction Imaging (T-CXDI). XPCS probes fluctuations in a speckle pattern by measuring the intensity-intensity (second order) correlation function as a function of the momentum transfer or scattering vector, q, $g^{(2)}(q,\tau)$

$$g^{(2)}(q,\tau) = \left\langle \frac{\langle I(q,t)I(q,t+\tau)\rangle_t}{\langle I(q,t)\rangle_t^2} \right\rangle \qquad 1)$$

where the brackets <> denote the ensemble averaging over all times t. For Gaussian fluctuations $g^{(2)}$ relates directly to the intermediate scattering function (ISF) $g^{(1)}$ or normalized dynamical structure factor

$$g^{(1)}(q,\tau) = \left|\frac{S(q,\tau)}{S(q,0)}\right| \qquad 2)$$

where S(q,0) is the static structure factor via the Siegert relation

$$g^{(2)}(q,\tau) = 1 + c\left|g^{(1)}(q,\tau)\right|^2 \qquad 3)$$

where c is the experimental contrast. The ISF probes the temporal relaxation of the Fourier component describing the dynamics of the system at the length scale 2π/q. $g^{(2)}$ is commonly measured by Dynamic Light Scattering (DLS) with fully coherent LASER sources providing a contrast often close to one. At visible wavelengths, DLS is applied to a complex system at the mesoscale (e.g., soft and hard colloidal suspensions, gels, and soft glasses [27]) for index-matched samples. On the other hand, X-rays are not restricted to index-matched samples and can look into dynamics at atomic length scales if a sufficient number of coherent X-ray photons are available.

Brilliance is the key parameter explaining the importance of sources like the ESRF EBS since for chaotic partially coherent sources like in synchrotron radiation storage rings, the coherent flux $F_c$ is limited to the emission over the diffraction limit,

$$F_c = B \left(\frac{\lambda}{2}\right)^2 \qquad 5)$$

at a wavelength $\lambda$ for a source of brilliance B, or photon flux per unit phase space volume (photons/s/mm$^2$/mr$^2$/0.1%band-width) [ref]. Having B as large as possible is fundamental to counteract the $\lambda^2$ decay, which is why coherent X-ray scattering is possible only at large synchrotron radiation facilities.

The suitable figure of merit explaining the importance of the brilliance for XPCS is the signal-to-noise ratio (SNR) of the g$^{(2)}$ of scattered intensities I(q)

$$SNR = \langle I(q)\rangle C\sqrt{\tau_{min} t N_{px}} \qquad 4)$$

where $\tau_{min}$ is the smallest time bin in the correlation function, t is the measuring time, $N_{px}$ is the number of pixels used for the ensemble averaging when a two-dimensional detector is used (i.e., $N_{px}$=1 for a point detector), and C is the contrast. Let us note that any increment in coherent flux brings a corresponding $\tau_{min}$ smaller by the square of the flux increment for the same SNR, i.e., an increase of the brilliance (coherent flux) by two orders of magnitude translates readily into an extension of the shortest time scales of four orders of magnitude for the same SNR.

Moreover, there is a substantial coherent fraction even at much higher energies than before, e.g., around 30 keV, opening experimental capabilities inaccessible so far, e.g., in bulky sample environments like diamond anvil cells for high-pressure research.

If we analyze how the increase of brilliance translates into improved coherent properties at ID10, the beam's coherence is retrieved via the statistical analysis of a random speckle pattern produced by a disordered system, a silica aerogel (Fig. 2A). In the condition of partial coherence, as is always the case for synchrotron radiation sources, the statistical properties of randomly scattered intensities (n) are well described by the distribution $P_M(n)$ of M-independent speckles patterns

$$P_M(n) = \left(\frac{M}{\langle n \rangle}\right)^M \frac{e^{\frac{-M\cdot n}{\langle n \rangle}} n^{M-1}}{\Gamma(M)} \qquad 5)$$

where $\Gamma$ is the gamma function. The distribution $P_M$(n) has mean <n> and variance <n>$^2$/M. In the first approximation, M can be considered as the number of coherence volumes in the scattering volume accounting for the observed contrast C=1/M.

This analysis has been applied to a random scattering pattern generated by a silica aerogel at small angles and 21.67 keV (Fig. 2B). The histograms of the probability events over a region with homogeneous count rates are well described by eq. 5) with M=8.6(2) coherent volumes. With a coherent flux of 6.5x10$^{11}$ photons/s over a focused-beam size of 4.9x3.9 µm$^2$ (HxV), the flux per coherent mode (volume) is 7.6x10$^{10}$ photons/s/mode. This number exceeds the coherent flux measured at the old high-β source but at the substantially lower energy of 8 keV.

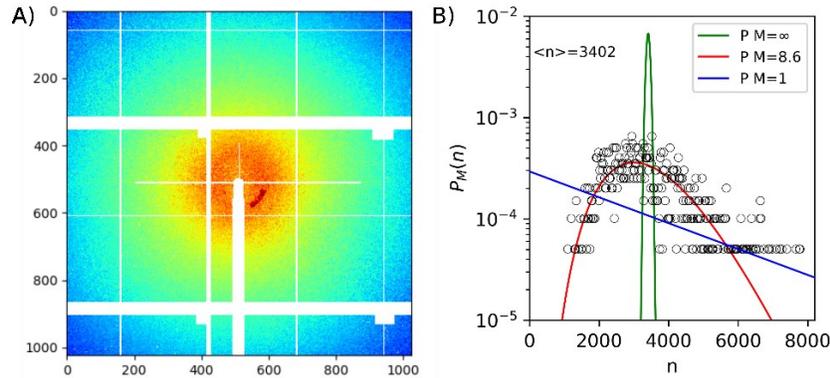

Fig. 2. The way how the increase of brilliance translates into improved coherent properties at ID10. The beam's coherence is retrieved via the statistical analysis of a random speckle pattern produced by a disordered system, a silica aerogel. A) small-angle X-ray scattering pattern from a silica aerogel under coherent illumination measured at 21.67 keV by a CdTe Eiger4M pixel detector located at 7.18 m from the sample. The shaded red sector has been used for the statistical analysis. B) the corresponding probability distribution $P_M(n)$, well described by eq. 5), in between the distribution for a perfectly coherent illumination (M=1, simple exponential) and the Poisson statistics in the incoherent limit (M=∞).

Another advantage of the higher natural collimation of the EBS beams is the availability of large horizontal transverse coherence lengths $\xi_T$

$$\xi_{T_{h,v}} = \frac{\lambda}{\pi} \frac{L}{S_{h,v}} \qquad 6)$$

where S is the source size and L is the source distance. For instance, at 60 m and 8.09 keV $\xi_{T,h}$ is about 45 μm. Fig. 3 shows the measured $g^{(2)}(q,\tau)$ from a dilute colloidal suspension of silica spheres in water, showing a contrast of about 7.6% with a 40 μm squared beam and 8.09 keV X-rays. Having the possibility to use larger beams is fundamental for radiation-sensitive (soft) systems as the absorbed dose can be substantially reduced for the same incident flux [28].

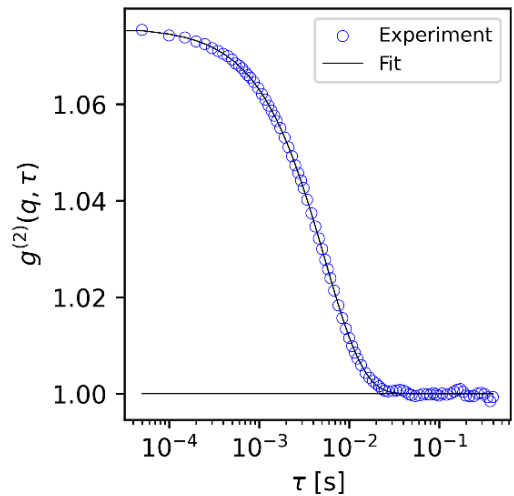

Fig. 3. The intensity autocorrelation function from a dilute colloidal suspension of silica spheres with a radius of 500 nm in water using a 40 μm squared beam at 8.09 keV. Data have been sampled at 20 kHz over 1s with an Eiger500K pixel detector and azimuthally averaged in an annulus of radius q=1.05x10$^{-2}$ nm$^{-1}$. The single exponential fit shows a contrast of about 7.6%.

At ID10, the coherent X-ray scattering activities are based on a four-circle diffractometer operating in horizontal scattering geometry and a sample-to-detector distance of up to 7 m. Depending on the desired geometry (small-angle (SA), wide-angle (WA), at grazing incidence), the flexibility of the setup allows to cover a q range from 10$^{-2}$ Å$^{-1}$ to 4 Å$^{-1}$. For WA-XPCS experiments, a second detector can be added close to the sample for simultaneous X-ray diffraction measurements, e.g., a Pilatus300K. The ID10 coherent X-ray scattering endstation is served by focusing optics based on Be compound-refractive lenses at 36 m and 53 m from the source capable of focusing X-rays from 7 keV to 16 keV with an energy step of roughly 1 keV and for 21 keV.

Besides XPCS, ID10 coherent X-ray scattering endstation host an instrument for CXDI for high-resolution imaging in three dimensions. CXDI is a lens-less, high-resolution imaging technique capable of reconstructing the electron density distribution at the resolution corresponding to the maximum q, where speckles are measured with a sufficient SNR. If the speckle pattern is oversampled better than the Nyquist rate $2^{1/N}$, where N is the space dimension, the speckle pattern can be phased via a numerical algorithm [29], and a simple Fourier transform retrieves the electron density. In the case of T-CXDI, the three-dimensional Fourier space is sampled by a tomographic θ scan of the sample over π (ideally), and the volumetric density is obtained without the need to assemble separate two-dimensional projections like in conventional tomography in real space. On the other hand, T-CXDI needs isolated samples and has a limited field of view. Since bigger objects produce smaller speckles, the field of view will be limited according to the oversampling conditions imposed by the detector pixel size with respect to the available detector distance and the wavelength [30]. In practice, at the ID10 coherent X-ray scattering endstation, we can image objects as large as 6 μm as the upper limit, ultimately limited by the maximum sample-to-detector distance (Fig. 4). The substantial gain in coherent flux and availability of a large detector (Eiger2 4M CdTe) allows measurements at higher scattering vectors and retrieving images at a resolution element of ~10 nm.

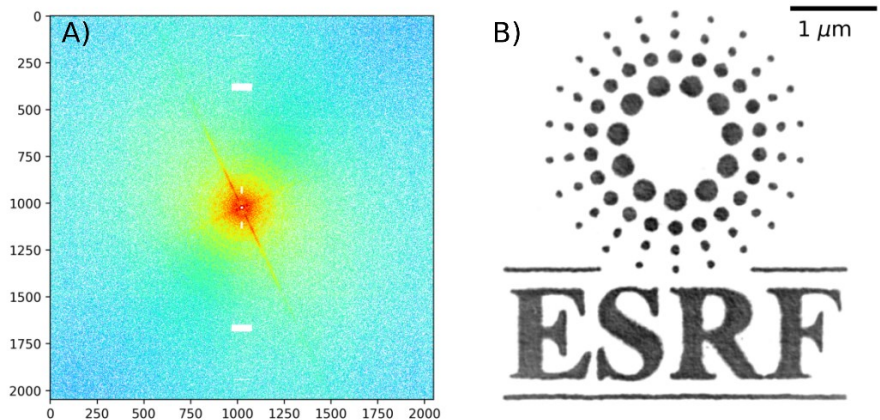

Fig. 4. An example of reconstruction obtained with EBS. The coherent diffraction pattern (left) from a test sample measured at 7.1 keV using Eiger2 4M CdTe detector. Reconstructed image of the test sample (right) with a pixel size of 7.6 nm.

## 2.2. ID10 surface scattering station

The ID10 surface scattering station is dedicated to studies of hard and soft condensed matter surfaces and interfaces using XRR, grazing incidence WAXS and SAXS, and auxiliary methods like X-ray fluorescence (XRF), X-ray absorption near edge structure (XANES), and surface XPCS [3]. The usable X-ray energy range is between 7-30 keV, and the beam is focused with a set of refractive lenses to the size of 10 x 25 µm at the sample spot. The sample is mounted on the 6+2 diffractometer, equipped with vertical and horizontal stages and a double crystal deflector (DCD) [31]. This device is used for the studies of liquid surfaces where the sample tilting is impossible. The main gains from EBS and beamline upgrade for the endstation are a much smaller beam in the horizontal direction, four times higher photon flux above 20 keV energy, and four orders of magnitude increase in coherent flux. These gains allow us to study 'photon hungry' samples, i.e., ultrathin-films [19] and high energy grazing incidence surface XPCS [32].

The studies of liquid surfaces and interfaces demand unique experimental geometry, i.e., the use of DCD, sample antivibration stage, and dedicated sample environments like the Langmuir trough. The routinely studied systems are two-dimensional crystals [7], membranes [33,34], and proteins [35] dispersed on the liquid surface, whose structure and order can be controlled by surface pressure variations in the Langmuir trough. In addition, the XANES experiments on protein layers can be carried out, as demonstrated by a recent study of the parkin protein adsorbed on a liquid interface on Langmuir trough [36], or standing wave studies on biological systems, i.e., pulmonary adsorption and biodistribution of gold nanoparticles [37]. Another studied class of materials are thin layers or nanosystems supported on solid substrates. The current investigations focus on the structure of perovskites [17,18,38], thin organic films [16,39,40], and polymers deposited on the nanopatterned surfaces [41]. Finally, a separate class of ID10 activities covers the studies of 2D materials primarily supported on liquid interfaces [19,42]. Dedicated sample environments and methodologies are necessary for cases where the synthesis conditions are harsh and hardly accessible for experimental methods other than X-ray-based [43,44].

The full exploitation of the "exotic" properties of the 2D materials demands the development of new synthesis methods that can lead to the growth of high-quality and large-area 2D crystals. For example,

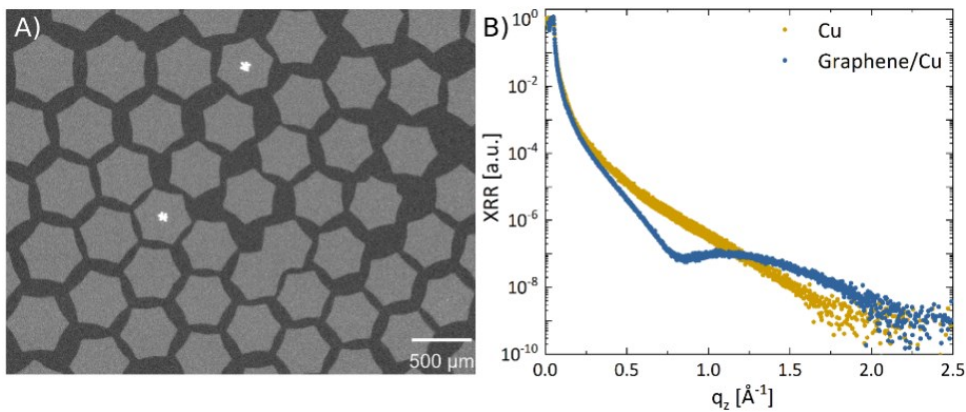

Fig. 5. Studies of graphene crystals floating on the surface of a liquid copper *in situ* and real-time. A) The radiation-mode optical microscopy image showing graphene crystals floating on the surface of liquid copper at 1400 K. The image was recorded during the graphene growth in the dedicated CVD reactor available at ID10. B) The XRR curves were recorded at ID10 using DCD and an X-ray beam of 22 keV. The orange curve was recorded on the bare surface of liquid copper, and the blue curve was recorded on the copper surface covered by graphene.

Graphene is a prototypical 2D material commonly grown on solid substrates by chemical vapor deposition (CVD) [45]. However, the resulting quality of this basic 2D material is far from perfect, limiting its broader applications. An alternative is CVD growth on liquid metal catalysts (LMCats) [46,47]. Unlike solids, these surfaces are isotropic, atomically smooth, without defects, and highly active catalytically [48]. These factors facilitate a much lower nucleation rate of the Graphene crystals, higher growth speed, and increased mobility on the surface, resulting in the growth of very high-quality Graphene. Recent developments implemented at ID10 allowed studies of Graphene crystals floating on the surface of LMCats *in situ* and in real-time. For this purpose, the prototype CVD reactor [44] was successfully commissioned at the ID10 beamline to perform X-ray diffraction experiments [43], complemented by radiation-mode optical microscopy [49] (Fig. 5A) and Raman spectroscopy. The possibility of real-time monitoring allowed us to gain control over Graphene growth and characterize its structure on the liquid copper surface using XRR and WAXS [19] (Fig. 5B). Consequently, it was concluded that even with low-intensity diffraction signal, it is possible to investigate the structure of 2D materials on liquid metals using synchrotron X-ray diffraction. These results were used to validate calculations based on first-principle theoretical methods [50]. The described experimental setup and the dedicated Raman spectrometer for high-temperature measurements are available for all the user community at the ID10 beamline.

## 3. ID02 Time-Resolved Ultra Small-Angle X-ray Scattering Beamline

The Time-Resolved Ultra SAXS (TRUSAXS) beamline ID02 is a multipurpose X-ray scattering instrument. The brilliance of the source is exploited for performing high angular and time resolution SAXS and related methods such as ultra SAXS (USAXS) and WAXS. These techniques have been widely used in the investigation of soft matter [3,13,51,52] and non-crystalline biological systems [9–12,53]. A unique feature is that these methods simultaneously offer a high spatial and temporal resolution, albeit in reciprocal space. Furthermore, scattering experiments can be combined with various thermophysical, rheological, and biophysical techniques (sample environments), further enhancing the structural and dynamical information [3,8,52]. New data analysis and computer simulation methods that emerged over the last decades have significantly increased the capacity of scattering methods to probe the complex soft matter and biological systems [9,13,53].

The main technical feature of the TRUSAXS instrument is an evacuated detector flight tube of 2 m diameter and 34 m length that enables continuous variation of sample-detector distance from about 0.8 m to 31 m [54]. Several high-performance detectors (Dectris Eiger2 4M, PSI Eiger2 500K, etc.) are housed inside a motorized wagon that travels along a rail system to the selected sample-detector distance. In addition, a wide-angle detector is appended to the SAXS detector tube. The combined scattering vector (of magnitude $q$) range covered is about $10^{-3}$ nm$^{-1}$ ≤ $q$ ≤ 50 nm$^{-1}$ for an X-ray wavelength of 1 Å. As a result, the beamline enables static and kinetic investigations of a broad range of systems from Å to µm size scales and down to sub-millisecond time range by combining different SAXS techniques in a single instrument.

Some representative scientific studies which exploit the instrument include probing the pathways of self-assembly in amphiphilic systems [3], interpolyelectrolyte complexes [51], detergent-mediated protein unfolding and refolding [13], virus capsid assembly [9,10], etc. The elucidation of transient structures formed by soft matter systems driven out-of-equilibrium constitutes another area of research [3,52]. The structural dynamics underlying the physiological activation of skeletal and cardiac muscle is a key topic that exploits the features of the beamline [11,12]. The accessible size and time scales enable an investigation of subtle changes induced to the ultrastructure of cells (e.g., by antimicrobial peptides) [53]. In addition, the beamline is also used for industrial research and development [22].

Additionally, a nearly coherent beam is obtained in the high-resolution mode allowing to perform multispeckle XPCS measurements down to the microsecond range over the ultra-small and small-angle regions [55]. An example is the recent work on the emergence of active dynamics of photocatalytic Janus colloids in $H_2O_2$ solution upon UV illumination [56]. In this case, XPCS enabled a separation of propulsive (mean velocity, velocity fluctuations) and diffusive (effective diffusion coefficient) components in the dynamics over a broad range of control parameters. The EBS permits to relax the collimation conditions, thereby obtaining a higher flux throughput and lower background. In particular, a coherent photon flux in excess of $2 \times 10^{12}$ photons/s can routinely be obtained in a 30 µm beam, allowing dynamical studies on relatively dilute samples.

Advanced pixel array detectors and high throughput data reduction pipelines complement the enhanced beam properties [54]. Together, these developments open new opportunities for structural, dynamic, and kinetic investigations of out-of-equilibrium soft matter [13,51,52,56] and biophysical systems [9–12,53]. Further exploitation of the EBS beam properties for SAXS/USAXS and XPCS methods is possible by utilizing a new focusing scheme based on compound refractive lenses. Both USAXS and UA-XPCS methods benefit from this development, enabling access to an unexplored range of low $q$ values. Fig. 6 summarizes the size and time scales accessible by scattering techniques for a sample that has appropriate structural features and electronic contrast.

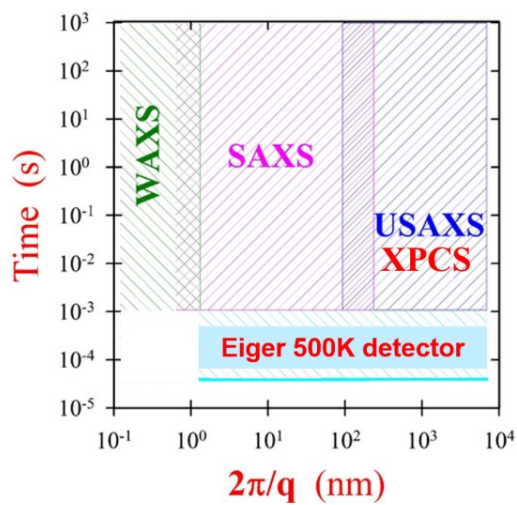

Fig. 6. The size and time scale accessible by the scattering methods available at ID02 for a system possessing adequate structural features over these ranges.

## 4. ID09 Pink Beam Time-Resolved Beamline

The ID09 is a beamline fully dedicated to time-resolved X-ray scattering/diffraction experiments [5,57]. The goal is to be able to follow structural changes taking place in a sample as a function of time after a given reaction-triggering event (Fig. 7, [58]). Phenomena that are routinely investigated at ID09 are chemical [59,60] or biological reactions [61,62] involving light-sensitive molecules dissolved in solution, light-induced phase transitions in solid-state samples [21,63], or laser-induced structural changes in colloidal systems [64]. These phenomena can be tracked as a function of time with 100 ps resolution up to seconds (or more). The beamline is equipped with several ultrafast laser systems able to provide laser pulses in a wide spectral range (from near-UV to near-IR) with picosecond or nanosecond duration. The high time resolution of 100 ps is achieved by isolating single X-ray pulses from the train generated by the ESRF source through an advanced system of fast choppers and shutters [65]. The pulse isolation results in a significant reduction of the X-ray average flux. To compensate for it, the beamline is designed in order to be able to operate without any monochromatizing optical element. This allows a very high number of photons ($10^8$-$10^9$) per pulse, which is the key parameter for time-resolved pump-probe experiments.

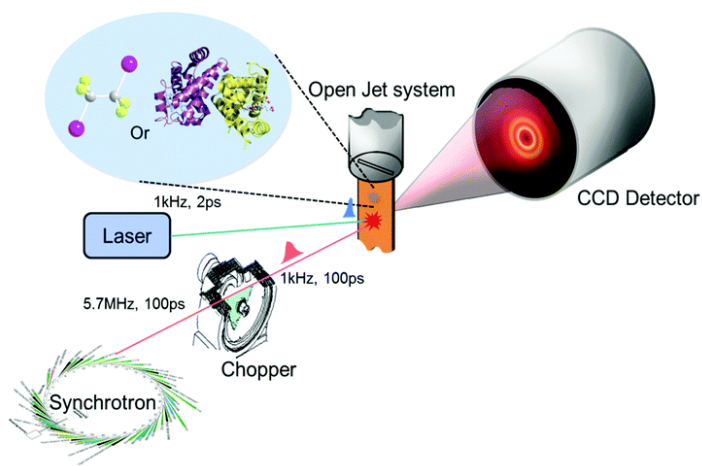

Fig. 7. Schematic representation of the setup for time-resolved X-ray solution scattering experiments at ID09. Single pink beam 100 ps X-ray pulses from the ESRF synchrotron are isolated using a high-speed mechanical chopper. These pulses are used to probe the structural changes induced by the laser pulses exciting the solutes in the sample. The laser-induced chemical or biological reactions are tracked by monitoring the time-resolved changes of the X-ray solution scattering patterns recorded by a CCD detector (figure taken from [58]).

ID09 has greatly benefited from the enhanced spectral purity of the beam resulting from the EBS upgrade. The main advantages are (1) a significant reduction of the second harmonic contamination, (2) bandwidth reduction, and (3) higher symmetry of the peak shape (Fig. 8). Thanks to these improved performances, time-resolved X-ray powder diffraction experiments [63,66] and serial crystallography experiments [67] can now be performed without employing any multilayer monochromator. Indeed, by combining the higher spectral purity of the EBS beam with a focusing mirror having a cutoff at ~24 keV, it is now possible to obtain a nearly symmetric first harmonic peak at the sample position with a ~1% bandwidth (Fig. 8, inset). This corresponds to an overall 5-fold increase in flux with respect to the pre-EBS experiments performed with a multilayer monochromator. Ultrafast pump-probe experiments, which are notoriously photon starving, can thus be performed with a 5-fold higher flux and without compromising the spectral purity necessary for extracting high-resolution information from the data.

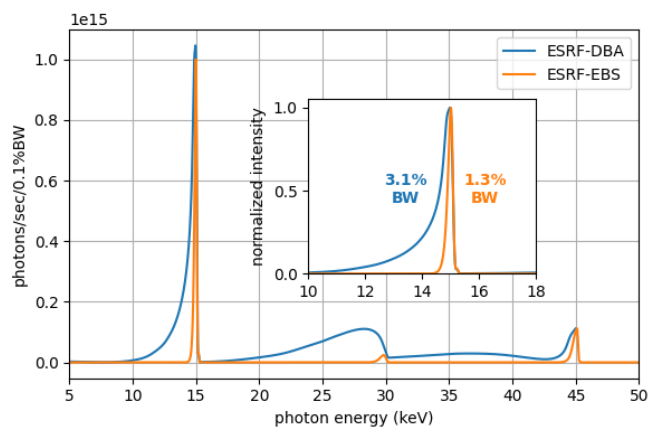

Fig. 8. Comparison of the X-ray beam spectrum before and after the EBS upgrade. Spectra produced by the in-vacuum U17 undulator available at the ID09 beamline in the typical beamline configuration used with the ESRF double-bend achromat (DBA) lattice (blue curve) and with the EBS MBA lattice (orange curve). The inset shows that the FWHM of the first harmonic peak is 2.4-fold narrower, and the peak shape is much more symmetric (full width at 10% of the maximum is 4.5-fold narrower).

## 5. Conclusions

The CBS group comprises three versatile beamlines for studies of soft and hard condensed matter using X-ray scattering, diffraction, and time-resolved methods. The optics design of three undulator beamlines. i.e., ID10, ID09, and ID02 guarantee enough beam brilliance for the X-ray photon-demanding investigations, including time-resolved and pump-probe experiments. The beamlines support multiple sample environments, working at different scattering geometries, for the *in situ* and *ex situ* studies. The ESRF-EBS upgrade has significantly improved the parameters of the generated X-ray beams, allowing for new types of experiments and studies of hitherto unexplored problems using synchrotron radiation.

## Acknowledgments

We thank the ESRF for providing synchrotron beam time and financial support. In addition, we thank all beamline personnel and ESRF support staff for contributing to the operation of the beamlines.

## References

[1] S.-M. Bak, Z. Shadike, R. Lin, X. Yu, X.-Q. Yang, In situ/operando synchrotron-based X-ray techniques for lithium-ion battery research, NPG Asia Mater. 10 (2018) 563–580. https://doi.org/10.1038/s41427-018-0056-z.

[2] P. Willmott, An Introduction to Synchrotron Radiation: Techniques and Applications, John Wiley & Sons, 2011.

[3] T. Narayanan, O. Konovalov, Synchrotron Scattering Methods for Nanomaterials and Soft Matter Research, Materials. 13 (2020) 752. https://doi.org/10.3390/ma13030752.

[4] F. Lehmkühler, W. Roseker, G. Grübel, From Femtoseconds to Hours—Measuring Dynamics over 18 Orders of Magnitude with Coherent X-rays, Appl. Sci. 11 (2021) 6179. https://doi.org/10.3390/app11136179.

[5] M. Wulff, A. Plech, L. Eybert, R. Randler, F. Schotte, P. Anfinrud, The realization of sub-nanosecond pump and probe experiments at the ESRF, Faraday Discuss. 122 (2003) 13–26. https://doi.org/10.1039/b202740m.

[6] P. Raimondi, ESRF-EBS: The Extremely Brilliant Source Project, Synchrotron Radiat. News. 29 (2016) 8–15. https://doi.org/10.1080/08940886.2016.1244462.

[7] J.J. Geuchies, C. van Overbeek, W.H. Evers, B. Goris, A. de Backer, A.P. Gantapara, F.T. Rabouw, J. Hilhorst, J.L. Peters, O. Konovalov, A.V. Petukhov, M. Dijkstra, L.D.A. Siebbeles, S. van Aert, S. Bals, D. Vanmaekelbergh, In situ study of the formation mechanism of two-dimensional superlattices from PbSe nanocrystals, Nat. Mater. 15 (2016) 1248–1254. https://doi.org/10.1038/nmat4746.

[8] M. Kapuscinski, M. Agthe, Z.-P. Lv, Y. Liu, M. Segad, L. Bergström, Temporal Evolution of Superlattice Contraction and Defect-Induced Strain Anisotropy in Mesocrystals during Nanocube Self-Assembly, ACS Nano. 14 (2020) 5337–5347. https://doi.org/10.1021/acsnano.9b07820.

[9] R. Asor, C.J. Schlicksup, Z. Zhao, A. Zlotnick, U. Raviv, Rapidly Forming Early Intermediate Structures Dictate the Pathway of Capsid Assembly, J. Am. Chem. Soc. 142 (2020) 7868–7882. https://doi.org/10.1021/jacs.0c01092.

[10] M. Chevreuil, L. Lecoq, S. Wang, L. Gargowitsch, N. Nhiri, E. Jacquet, T. Zinn, S. Fieulaine, S. Bressanelli, G. Tresset, Nonsymmetrical Dynamics of the HBV Capsid Assembly and Disassembly Evidenced by Their Transient Species, J. Phys. Chem. B. 124 (2020) 9987–9995. https://doi.org/10.1021/acs.jpcb.0c05024.

[11] M. Reconditi, M. Caremani, F. Pinzauti, J.D. Powers, T. Narayanan, G.J.M. Stienen, M. Linari, V. Lombardi, G. Piazzesi, Myosin filament activation in the heart is tuned to the mechanical task, Proc. Natl. Acad. Sci. 114 (2017) 3240–3245. https://doi.org/10.1073/pnas.1619484114.

[12] E. Brunello, L. Fusi, A. Ghisleni, S.-J. Park-Holohan, J.G. Ovejero, T. Narayanan, M. Irving, Myosin filament-based regulation of the dynamics of contraction in heart muscle, Proc. Natl. Acad. Sci. 117 (2020) 8177–8186. https://doi.org/10.1073/pnas.1920632117.

[13] J.N. Pedersen, J. Lyngsø, T. Zinn, D.E. Otzen, J.S. Pedersen, A complete picture of protein unfolding and refolding in surfactants, Chem. Sci. 11 (2020) 699–712. https://doi.org/10.1039/C9SC04831F.

[14] Y. Chushkin, A. Gulotta, F. Roosen-Runge, A. Pal, A. Stradner, P. Schurtenberger, Probing cage relaxation in concentrated protein solutions by XPCS, (2022). https://doi.org/10.48550/arXiv.2203.12695.

[15] T. Beuvier, I. Probert, L. Beaufort, B. Suchéras-Marx, Y. Chushkin, F. Zontone, A. Gibaud, X-ray nanotomography of coccolithophores reveals that coccolith mass and segment number correlate with grid size, Nat. Commun. 10 (2019) 751. https://doi.org/10.1038/s41467-019-08635-x.

[16] C. Wang, N. Russegger, G. Duva, O. V. Konovalov, M. Jankowski, A. Gerlach, A. Hinderhofer, F. Schreiber, Growth, structure and templating of anthradithiophene and its β-methylthiolated derivative, Mater. Chem. Front. (2022). https://doi.org/10.1039/D2QM00759B.

[17] T. Antrack, M. Kroll, L. Merten, M. Albaladejo-Siguan, J. Benduhn, A. Hinderhofer, O. Konovalov, M. Jankowski, F. Schreiber, Y. Vaynzof, K. Leo, Enhancing Luminescence Efficiency by Controlled Island Formation of $CsPbBr_3$ Perovskite, Adv. Opt. Mater. (2022).

[18] O. Telschow, M. Albaladejo-Siguan, L. Merten, A.D. Taylor, K.P. Goetz, T. Schramm, O.V. Konovalov, M. Jankowski, A. Hinderhofer, F. Paulus, F. Schreiber, Y. Vaynzof, Preserving the stoichiometry of triple-cation perovskites by carrier-gas-free antisolvent spraying, J. Mater. Chem. A. (2022). https://doi.org/10.1039/D1TA10566C.

[19] M. Jankowski, M. Saedi, F. La Porta, A.C. Manikas, C. Tsakonas, J.S. Cingolani, M. Andersen, M. de Voogd, G.J.C. van Baarle, K. Reuter, C. Galiotis, G. Renaud, O.V. Konovalov, I.M.N. Groot, Real-Time Multiscale Monitoring and Tailoring of Graphene Growth on Liquid Copper, ACS Nano. 15 (2021) 9638–9648. https://doi.org/10.1021/acsnano.0c10377.

[20] S. Thomet, F. Ghaffari, S. De Paoli, J.-M. Daveau, F. Abouzeid, O. Romain, Observation framework of errors in microprocessors with machine learning location inference of radiation-induced faults, Microelectron. Reliab. 137 (2022) 114667. https://doi.org/10.1016/j.microrel.2022.114667.

[21] A. Volte, C. Mariette, R. Bertoni, M. Cammarata, X. Dong, E. Trzop, H. Cailleau, E. Collet, M. Levantino, M. Wulff, J. Kubicki, F.-L. Yang, M.-L. Boillot, B. Corraze, L. Stoleriu, C. Enachescu, M. Lorenc, Dynamical limits for the molecular switching in a photoexcited material revealed by X-ray diffraction, Commun. Phys. 5 (2022) 1–9. https://doi.org/10.1038/s42005-022-00940-0.

[22] M. Staropoli, D. Gerstner, M. Sztucki, G. Vehres, B. Duez, S. Westermann, D. Lenoble, W. Pyckhout-Hintzen, Hierarchical Scattering Function for Silica-Filled Rubbers under Deformation: Effect of the Initial Cluster Distribution, Macromolecules. 52 (2019) 9735–9745. https://doi.org/10.1021/acs.macromol.9b01751.

[23] D. Chenevier, A. Joly, ESRF: Inside the Extremely Brilliant Source Upgrade, Synchrotron Radiat. News. 31 (2018) 32–35. https://doi.org/10.1080/08940886.2018.1409562.

[24] L. Leu, A. Georgiadis, M.J. Blunt, A. Busch, P. Bertier, K. Schweinar, M. Liebi, A. Menzel, H. Ott, Multiscale Description of Shale Pore Systems by Scanning SAXS and WAXS Microscopy, Energy Fuels. 30 (2016) 10282–10297. https://doi.org/10.1021/acs.energyfuels.6b02256.

[25] F. Zontone, A. Madsen, O. Konovalov, R. Garrett, I. Gentle, K. Nugent, S. Wilkins, Measuring The Source Brilliance at An Undulator Beamline, in: Melbourne (Australia), 2010: pp. 603–606. https://doi.org/10.1063/1.3463279.

[26] O. Chubar, P. Elleaume, Accurate and efficient computation of synchrotron radiation in the near field region, Conf Proc C. 980622 (1998) 1177–1179.

[27] E.R. Weeks, Introduction to the Colloidal Glass Transition, ACS Macro Lett. 6 (2017) 27–34. https://doi.org/10.1021/acsmacrolett.6b00826.

[28] A. Girelli, H. Rahmann, N. Begam, A. Ragulskaya, M. Reiser, S. Chandran, F. Westermeier, M. Sprung, F. Zhang, C. Gutt, F. Schreiber, Microscopic Dynamics of Liquid-Liquid Phase Separation and Domain

Coarsening in a Protein Solution Revealed by X-Ray Photon Correlation Spectroscopy, Phys. Rev. Lett. 126 (2021) 138004. https://doi.org/10.1103/PhysRevLett.126.138004.

[29] J.R. Fienup, Phase retrieval algorithms: a comparison, Appl. Opt. 21 (1982) 2758. https://doi.org/10.1364/AO.21.002758.

[30] J. Miao, D. Sayre, H.N. Chapman, Phase retrieval from the magnitude of the Fourier transforms of nonperiodic objects, J. Opt. Soc. Am. A. 15 (1998) 1662. https://doi.org/10.1364/JOSAA.15.001662.

[31] O. Konovalov, V. Belova, M. Saedi, I. Groot, G. Renaud, M. Jankowski, Tripling of the scattering vector range of X-ray reflectivity on liquid surfaces using a double crystal deflector, (2022). https://doi.org/10.48550/arXiv.2210.12827.

[32] F. Amadei, J. Thoma, J. Czajor, E. Kimmle, A. Yamamoto, W. Abuillan, O.V. Konovalov, Y. Chushkin, M. Tanaka, Ion-Mediated Cross-linking of Biopolymers Confined at Liquid/Liquid Interfaces Probed by In Situ High-Energy Grazing Incidence X-ray Photon Correlation Spectroscopy, J. Phys. Chem. B. 124 (2020) 8937–8942. https://doi.org/10.1021/acs.jpcb.0c07056.

[33] L.-G. Bronstein, P. Cressey, W. Abuillan, O. Konovalov, M. Jankowski, V. Rosilio, A. Makky, Influence of the porphyrin structure and linker length on the interfacial behavior of phospholipid-porphyrin conjugates, J. Colloid Interface Sci. 611 (2022) 441–450.

[34] J. Massiot, W. Abuillan, O. Konovalov, A. Makky, Photo-triggerable liposomes based on lipid-porphyrin conjugate and cholesterol combination: Formulation and mechanistic study on monolayers and bilayers, Biochim. Biophys. Acta BBA - Biomembr. 1864 (2022) 183812. https://doi.org/10.1016/j.bbamem.2021.183812.

[35] G. Surmeier, S. Dogan-Surmeier, M. Paulus, C. Albers, J. Latarius, C. Sternemann, E. Schneider, M. Tolan, J. Nase, The interaction of viral fusion peptides with lipid membranes, Biophys. J. 121 (2022) 3811–3825. https://doi.org/10.1016/j.bpj.2022.09.011.

[36] O.V. Konovalov, N.N. Novikova, M.V. Kovalchuk, G.E. Yalovega, A.F. Topunov, O.V. Kosmachevskaya, E.A. Yurieva, A.V. Rogachev, A.L. Trigub, M.A. Kremennaya, V.I. Borshchevskiy, D.D. Vakhrameev, S.N. Yakunin, XANES Measurements for Studies of Adsorbed Protein Layers at Liquid Interfaces, Materials. 13 (2020) 4635. https://doi.org/10.3390/ma13204635.

[37] A.V. Rogachev, N.N. Novikova, M.V. Kovalchuk, Yu.N. Malakhova, O.V. Konovalov, N.D. Stepina, E.A. Shlyapnikova, I.L. Kanev, Yu.M. Shlyapnikov, S.N. Yakunin, Permeation of Nanoparticles into Pulmonary Surfactant Monolayer: In Situ X-ray Standing Wave Studies, Langmuir. 38 (2022) 3630–3640. https://doi.org/10.1021/acs.langmuir.1c02179.

[38] K.O. Brinkmann, T. Becker, F. Zimmermann, C. Kreusel, T. Gahlmann, M. Theisen, T. Haeger, S. Olthof, C. Tückmantel, M. Günster, T. Maschwitz, F. Göbelsmann, C. Koch, D. Hertel, P. Caprioglio, F. Peña-Camargo, L. Perdigón-Toro, A. Al-Ashouri, L. Merten, A. Hinderhofer, L. Gomell, S. Zhang, F. Schreiber, S. Albrecht, K. Meerholz, D. Neher, M. Stolterfoht, T. Riedl, Perovskite–organic tandem solar cells with indium oxide interconnect, Nature. 604 (2022) 280–286. https://doi.org/10.1038/s41586-022-04455-0.

[39] V. Belova, B. Wagner, B. Reisz, C. Zeiser, G. Duva, J. Rozbořil, J. Novák, A. Gerlach, A. Hinderhofer, F. Schreiber, Real-Time Structural and Optical Study of Growth and Packing Behavior of Perylene Diimide Derivative Thin Films: Influence of Side-Chain Modification, J. Phys. Chem. C. 122 (2018) 8589–8601. https://doi.org/10.1021/acs.jpcc.8b00787.

[40] V. Belova, A. Hinderhofer, C. Zeiser, T. Storzer, J. Rozbořil, J. Hagenlocher, J. Novák, A. Gerlach, R. Scholz, F. Schreiber, Structure-Dependent Charge Transfer in Molecular Perylene-Based Donor/Acceptor Systems and Role of Side Chains, J. Phys. Chem. C. 124 (2020) 11639–11651. https://doi.org/10.1021/acs.jpcc.0c00230.

[41] R. Ruffino, L. Fichera, A. Valenti, M. Jankowski, O. Konovalov, G.M.L. Messina, A. Licciardello, N. Tuccitto, G. Li-Destri, G. Marletta, Tuning the randomization of lamellar orientation in poly(3-

hexylthiophene) thin films with substrate nano-curvature, Polymer. 230 (2021) 124071. https://doi.org/10.1016/j.polymer.2021.124071.

[42] P. Ravat, H. Uchida, R. Sekine, K. Kamei, A. Yamamoto, O. Konovalov, M. Tanaka, T. Yamada, K. Harano, E. Nakamura, De Novo Synthesis of Free-Standing Flexible 2D Intercalated Nanofilm Uniform over Tens of cm2, Adv. Mater. 34 (2022) 2106465. https://doi.org/10.1002/adma.202106465.

[43] O.V. Konovalov, V. Belova, F. La Porta, M. Saedi, I.M.N. Groot, G. Renaud, I. Snigireva, A. Snigirev, M. Voevodina, C. Shen, A. Sartori, B.M. Murphy, M. Jankowski, X-ray reflectivity from curved surfaces as illustrated by a graphene layer on molten copper, J. Synchrotron Radiat. 29 (2022). https://doi.org/10.1107/S1600577522002053.

[44] M. Saedi, J.M. de Voogd, A. Sjardin, A. Manikas, C. Galiotis, M. Jankowski, G. Renaud, F. La Porta, O. Konovalov, G.J.C. van Baarle, I.M.N. Groot, Development of a reactor for the *in situ* monitoring of 2D materials growth on liquid metal catalysts, using synchrotron X-ray scattering, Raman spectroscopy, and optical microscopy, Rev. Sci. Instrum. 91 (2020) 013907. https://doi.org/10.1063/1.5110656.

[45] M. Batzill, The surface science of graphene: Metal interfaces, CVD synthesis, nanoribbons, chemical modifications, and defects, Surf. Sci. Rep. 67 (2012) 83–115. https://doi.org/10.1016/j.surfrep.2011.12.001.

[46] D. Geng, B. Wu, Y. Guo, L. Huang, Y. Xue, J. Chen, G. Yu, L. Jiang, W. Hu, Y. Liu, Uniform hexagonal graphene flakes and films grown on liquid copper surface, Proc. Natl. Acad. Sci. 109 (2012) 7992–7996. https://doi.org/10.1073/pnas.1200339109.

[47] C. Tsakonas, M. Dimitropoulos, A.C. Manikas, C. Galiotis, Growth and *in situ* characterization of 2D materials by chemical vapour deposition on liquid metal catalysts: a review, Nanoscale. 13 (2021) 3346–3373. https://doi.org/10.1039/D0NR07330J.

[48] P. Aukarasereenont, A. Goff, C.K. Nguyen, C.F. McConville, A. Elbourne, A. Zavabeti, T. Daeneke, Liquid metals: an ideal platform for the synthesis of two-dimensional materials, Chem. Soc. Rev. (2022). https://doi.org/10.1039/D1CS01166A.

[49] T. Terasawa, K. Saiki, Radiation-mode optical microscopy on the growth of graphene, Nat. Commun. 6 (2015) 6834. https://doi.org/10.1038/ncomms7834.

[50] H. Gao, V. Belova, F. La Porta, J.S. Cingolani, M. Andersen, M. Saedi, O.V. Konovalov, M. Jankowski, H.H. Heenen, I.M.N. Groot, G. Renaud, K. Reuter, Graphene at Liquid Copper Catalysts: Atomic-Scale Agreement of Experimental and First-Principles Adsorption Height, Adv. Sci. n/a (2022) 2204684. https://doi.org/10.1002/advs.202204684.

[51] R. Takahashi, T. Narayanan, S. Yusa, T. Sato, Formation Kinetics of Polymer Vesicles from Spherical and Cylindrical Micelles Bearing the Polyelectrolyte Complex Core Studied by Time-Resolved USAXS and SAXS, Macromolecules. 55 (2022) 684–695. https://doi.org/10.1021/acs.macromol.1c02210.

[52] L. Matthews, T. Narayanan, High-resolution structural elucidation of extremely swollen lyotropic phases, J. Colloid Interface Sci. 610 (2022) 359–367. https://doi.org/10.1016/j.jcis.2021.11.168.

[53] E.F. Semeraro, L. Marx, J. Mandl, I. Letofsky-Papst, C. Mayrhofer, M.P. Frewein, H.L. Scott, S. Prévost, H. Bergler, K. Lohner, G. Pabst, Lactoferricins impair the cytosolic membrane of Escherichia coli within a few seconds and accumulate inside the cell, ELife. 11 (2022) e72850. https://doi.org/10.7554/eLife.72850.

[54] T. Narayanan, M. Sztucki, T. Zinn, J. Kieffer, A. Homs-Puron, J. Gorini, P. Van Vaerenbergh, P. Boesecke, Performance of the time-resolved ultra-small-angle X-ray scattering beamline with the Extremely Brilliant Source, J. Appl. Crystallogr. 55 (2022) 98–111. https://doi.org/10.1107/S1600576721012693.

[55] T. Zinn, L. Sharpnack, T. Narayanan, Phoretic dynamics of colloids in a phase separating critical liquid mixture, Phys. Rev. Res. 2 (2020) 033177. https://doi.org/10.1103/PhysRevResearch.2.033177.

[56] T. Zinn, T. Narayanan, S.N. Kottapalli, J. Sachs, T. Sottmann, P. Fischer, Emergent dynamics of light-induced active colloids probed by XPCS, New J. Phys. 24 (2022) 093007. https://doi.org/10.1088/1367-2630/ac8a66.

[57] M. Levantino, Q. Kong, M. Cammarata, D. Khakhulin, F. Schotte, P. Anfinrud, V. Kabanova, H. Ihee, A. Plech, S. Bratos, M. Wulff, Structural dynamics probed by X-ray pulses from synchrotrons and XFELs, Comptes Rendus Phys. 22 (2021) 75–94. https://doi.org/10.5802/crphys.85.

[58] J. Kim, K.H. Kim, K.Y. Oang, J.H. Lee, K. Hong, H. Cho, N. Huse, R.W. Schoenlein, T.K. Kim, H. Ihee, Tracking reaction dynamics in solution by pump–probe X-ray absorption spectroscopy and X-ray liquidography (solution scattering), Chem. Commun. 52 (2016) 3734–3749. https://doi.org/10.1039/C5CC08949B.

[59] G. Smolentsev, C.J. Milne, A. Guda, K. Haldrup, J. Szlachetko, N. Azzaroli, C. Cirelli, G. Knopp, R. Bohinc, S. Menzi, G. Pamfilidis, D. Gashi, M. Beck, A. Mozzanica, D. James, C. Bacellar, G.F. Mancini, A. Tereshchenko, V. Shapovalov, W.M. Kwiatek, J. Czapla-Masztafiak, A. Cannizzo, M. Gazzetto, M. Sander, M. Levantino, V. Kabanova, E. Rychagova, S. Ketkov, M. Olaru, J. Beckmann, M. Vogt, Taking a snapshot of the triplet excited state of an OLED organometallic luminophore using X-rays, Nat. Commun. 11 (2020) 2131. https://doi.org/10.1038/s41467-020-15998-z.

[60] H. Ki, T.W. Kim, J. Moon, J. Kim, Y. Lee, J. Heo, K.H. Kim, Q. Kong, D. Khakhulin, G. Newby, J. Kim, J. Kim, M. Wulff, H. Ihee, Photoactivation of triosmium dodecacarbonyl at 400 nm probed with time-resolved X-ray liquidography, Chem. Commun. 58 (2022) 7380–7383. https://doi.org/10.1039/D2CC02438A.

[61] F. Orädd, H. Ravishankar, J. Goodman, P. Rogne, L. Backman, A. Duelli, M. Nors Pedersen, M. Levantino, M. Wulff, M. Wolf-Watz, M. Andersson, Tracking the ATP-binding response in adenylate kinase in real time, Sci. Adv. 7 (2021) eabi5514. https://doi.org/10.1126/sciadv.abi5514.

[62] D. Sarabi, L. Ostojić, R. Bosman, A. Vallejos, J.-B. Linse, M. Wulff, M. Levantino, R. Neutze, Modeling difference x-ray scattering observations from an integral membrane protein within a detergent micelle, Struct. Dyn. 9 (2022) 054102. https://doi.org/10.1063/4.0000157.

[63] C. Mariette, M. Lorenc, H. Cailleau, E. Collet, L. Guérin, A. Volte, E. Trzop, R. Bertoni, X. Dong, B. Lépine, O. Hernandez, E. Janod, L. Cario, V. Ta Phuoc, S. Ohkoshi, H. Tokoro, L. Patthey, A. Babic, I. Usov, D. Ozerov, L. Sala, S. Ebner, P. Böhler, A. Keller, A. Oggenfuss, T. Zmofing, S. Redford, S. Vetter, R. Follath, P. Juranic, A. Schreiber, P. Beaud, V. Esposito, Y. Deng, G. Ingold, M. Chergui, G.F. Mancini, R. Mankowsky, C. Svetina, S. Zerdane, A. Mozzanica, A. Bosak, M. Wulff, M. Levantino, H. Lemke, M. Cammarata, Strain wave pathway to semiconductor-to-metal transition revealed by time-resolved X-ray powder diffraction, Nat. Commun. 12 (2021) 1239. https://doi.org/10.1038/s41467-021-21316-y.

[64] A. Plech, A.R. Ziefuß, M. Levantino, R. Streubel, S. Reich, S. Reichenberger, Low Efficiency of Laser Heating of Gold Particles at the Plasmon Resonance: An X-ray Calorimetry Study, ACS Photonics. 9 (2022) 2981–2990. https://doi.org/10.1021/acsphotonics.2c00588.

[65] M. Cammarata, L. Eybert, F. Ewald, W. Reichenbach, M. Wulff, P. Anfinrud, F. Schotte, A. Plech, Q. Kong, M. Lorenc, B. Lindenau, J. Räbiger, S. Polachowski, Chopper system for time resolved experiments with synchrotron radiation, Rev. Sci. Instrum. 80 (2009) 015101. https://doi.org/10.1063/1.3036983.

[66] C.M. Pépin, A. Sollier, A. Marizy, F. Occelli, M. Sander, R. Torchio, P. Loubeyre, Kinetics and structural changes in dynamically compressed bismuth, Phys. Rev. B. 100 (2019) 060101. https://doi.org/10.1103/PhysRevB.100.060101.

[67] A. Tolstikova, M. Levantino, O. Yefanov, V. Hennicke, P. Fischer, J. Meyer, A. Mozzanica, S. Redford, E. Crosas, N.L. Opara, M. Barthelmess, J. Lieske, D. Oberthuer, E. Wator, I. Mohacsi, M. Wulff, B. Schmitt, H.N. Chapman, A. Meents, 1 kHz fixed-target serial crystallography using a multilayer

monochromator and an integrating pixel detector, IUCrJ. 6 (2019) 927–937. https://doi.org/10.1107/S205225251900914X.